# Age dependence of fitness and body mass index in Korean adults

Nam Lyong Kang*, Su Chak Ryu

Department of Nanomechatronics Engineering, Pusan National University, Miryang 50463, Republic of Korea

*Corresponding author: Department of Nanomechatronics Engineering, 1268-50 Samnangjin-ro, Samnangjin-eup, Miryang-si, Gyeonsangnam-do 50463, Republic of Korea
E-mail: nlkang@pusan.ac.kr

## Abstract

The aim of this study was to investigate the age dependence of the fitness and body mass index (BMI) in Korean adults and to find an effective exercise to restore the degradation of fitness due to aging. The age dependence of the fitness and BMI were calculated using their lump mean values (LMVs) and a linear regression method. The fitness sensitivity percentage to age (FSPA) and fitness sensitivity percentage to BMI (FSPB) were introduced as indicators for the effective improvement of the fitness. The results showed that the degradation of fitness due to aging, especially the degradation of cardiorespiratory endurance and muscular endurance, could be improved effectively by controlling the 20-m multi-stage shuttle run and sit-up scores for both males and females. The results also showed that the BMIs could be effectively controlled with enhancing the 10-m shuttle run and standing long jump scores for both males and females. It is expected that the LMV, FSPA, and FSPB could be used to improve fitness effectively and to establish personal exercise aims.

# 1. Introduction

With the increase of mortality related to abdominal obesity, an exact assessment of fitness is more important for reducing them through physical activity or exercise. Abdominal obesity is known as a major risk factor for diseases such as diabetes and hypertension, but it can be reduced by physical activity (Dubbert et al., 2002; Lakka et al., 1994; Lavie et al., 2019; Matsuzawa, 2010). Therefore, an effective index is needed for the enhancement of fitness and for use as an effective means of preventing and treating abdominal obesity.

The body mass index (BMI) has been considered as a useful index for assessing obesity because it is not intrusive and is easy to calculate with acceptable accuracy (Hu et al., 2004; Hwang et al., 2020; Kawai et al., 2019; Pasco et al., 2014; Raustorp et al., 2004). But BMI is not a proper index for measuring the abdominal obesity because it includes both fat and muscle. Thus, the waist circumference has been considered as an alternative anthropometric index that is more convenient and less expensive (Ashwell et al., 2012; Hsieh et al., 2003; Krakauer et al., 2012; Nevill et al., 2017; Seidell, 2010; Shen et al., 2017; Taylor et al., 1998; Vasquez et al., 2019; Zhu et al., 2002). However, BMI is a more proper index than indices including waist circumference for assessing fitness metrics such as muscular endurance, quickness, cardiorespiratory endurance, speed, and agility because waist circumference is affected mostly by fat and is nearly unrelated to muscle.

In general, fitness decreases and BMI increases as humans age, and the degradation differs according to the type of fitness. Therefore, it is necessary to investigate the age dependence of the relation between fitness and BMI. This paper introduces the fitness sensitivity percentage to age (FSPA) and fitness sensitivity percentage to BMI (FSPB) to investigate useful exercises that enhance fitness that has been weakened by aging.

On the order hand, it is nearly impossible to investigate the age dependence of BMI and fitness individually because they vary with age in a very complicated manner. Thus, this paper investigates them using the lump mean value (LMV) (Hwang et al., 2020) and a linear regression method, which fits the LMVs of each fitness test score and BMI with respect to age to a straight line. The age dependence of fitness and BMI is obtained from the slope of the straight line, and the FSPA and FSPB are also determined by the slope and the best value of each fitness test.

## 2. Methods

*2.1. Study design and participants*

This cross-sectional study analyzed the public-use releases of the 2017 Survey of National Physical Fitness (ISBN 979-11-952035-6-7) by the Korea Institute of Sport Science of the Korea Sports Promotion Foundation (KSPO). The survey is a nationally representative cross-sectional survey. The collection of data and samples from the cohorts participating in the study was approved by Statistics Korea (National Statistics No 113004). Furthermore, all research participants provided written consent to participate. A detailed description of the plan and the operation of the survey are available on the Korea Institute of Sport Science website (http://www.sports.re.kr, available in Korean).

The four fitness tests considered were sit-up, standing long jump (SLJ), 20-m multi-stage shuttle run (20-m MSSR), and 10-m shuttle run (10-m SR) tests (Hwang et al., 2020). The sit-up test is for measuring muscular endurance, the SLJ test is for quickness, the 20-m MSSR test is for cardiorespiratory endurance, and the 10-m SR test is for speed and agility. The final study cohort comprised 1902 males and 1861 females from a total of 3763 subjects aged 20 to 59 years. Subjects were excluded if they had extreme values of BMI or scores on the four fitness tests. The following ranges were considered: $17.96 \leq \text{BMI (kg/m}^2) \leq 35.67$, $2 \leq \text{sit-up (times/minute)} \leq 72$, $61 \leq \text{SLJ (cm)} \leq 280$, $3 \leq \text{20-m MSSR (times)} \leq 100$, and $8.5 \leq \text{10-m SR (seconds)} \leq 19.31$ for males and $15.98 \leq \text{BMI (kg/m}^2) \leq 34.87$, $2 \leq \text{sit-up (times/minute)} \leq 65$, $50 \leq \text{SLJ (cm)} \leq 223$, $2 \leq \text{20-m MSSR (times)} \leq 82$, and $9.5 \leq \text{10-m SR (seconds)} \leq 21.43$ for females.

*2.2. Data analysis*

The LMVs (Tables 1 and 2) of fitness scores and BMI were considered because their distributions with respect to age are very complicated. Each lump was composed of subjects with two successive ages, and the number of subjects contained in each lump was slightly different according to age. This study investigated the age dependence of BMI and fitness using the Microsoft Excel 2014 and a linear regression in Sigmaplot 14.

**Table 1**

Lump mean values of BMI (kg/m$^2$), sit-up score (times/minute), SLJ score (cm), 20-m MSSR score (times), and 10-m SR score (second) for males.

| Age | N | BMI | | sit-up | | SLJ | | 20-m MSSR | | 10-m SR | |
|---|---|---|---|---|---|---|---|---|---|---|---|
| | | Mean | SD | Mean | SD | Mean | SD | Mean | SD | Mean | SD |
| 20.5 | 65 | 23.3 | 2.53 | 49 | 12.8 | 224 | 27.4 | 53 | 18.1 | 10.3 | 0.96 |
| 22.5 | 114 | 24.1 | 3.01 | 48 | 11.8 | 218 | 27.7 | 51 | 18.9 | 10.7 | 1.25 |
| 24.5 | 105 | 23.7 | 2.59 | 47 | 9.83 | 218 | 25.3 | 48 | 19.5 | 10.8 | 1.28 |
| 26.5 | 92 | 24.5 | 2.74 | 45 | 10.6 | 210 | 24.0 | 46 | 18.4 | 10.7 | 1.10 |
| 28.5 | 67 | 24.9 | 3.48 | 45 | 12.2 | 213 | 21.6 | 43 | 19.9 | 10.9 | 1.17 |
| 30.5 | 98 | 25.1 | 2.84 | 44 | 11.6 | 213 | 25.6 | 44 | 18.3 | 10.9 | 1.48 |
| 32.5 | 104 | 25.7 | 3.22 | 42 | 12.5 | 207 | 28.7 | 43 | 17.8 | 11.1 | 1.30 |
| 34.5 | 96 | 24.7 | 2.71 | 38 | 11.4 | 205 | 25.9 | 39 | 16.7 | 11.5 | 1.39 |
| 36.5 | 96 | 24.8 | 2.82 | 42 | 10.9 | 205 | 22.6 | 41 | 16.9 | 11.2 | 1.08 |
| 38.5 | 84 | 25.2 | 2.79 | 39 | 11.1 | 204 | 28.1 | 38 | 15.0 | 11.5 | 1.57 |
| 40.5 | 87 | 25.7 | 2.95 | 40 | 8.80 | 201 | 24.0 | 37 | 15.1 | 11.4 | 1.13 |
| 42.5 | 105 | 25.1 | 3.02 | 38 | 11.6 | 198 | 22.7 | 35 | 14.5 | 11.7 | 1.06 |
| 44.5 | 86 | 25.3 | 3.11 | 38 | 11.2 | 199 | 25.4 | 37 | 18.1 | 11.6 | 1.41 |
| 46.5 | 112 | 25.2 | 2.93 | 37 | 9.14 | 195 | 22.9 | 33 | 13.2 | 11.9 | 1.12 |
| 48.5 | 92 | 25.1 | 2.85 | 36 | 10.5 | 190 | 20.6 | 33 | 13.2 | 12.1 | 1.54 |
| 50.5 | 96 | 24.5 | 2.68 | 34 | 10.3 | 189 | 22.0 | 32 | 14.1 | 12.1 | 1.39 |
| 52.5 | 108 | 24.8 | 2.87 | 33 | 10.6 | 185 | 19.3 | 31 | 14.6 | 12.3 | 1.29 |
| 54.5 | 95 | 24.9 | 2.64 | 32 | 10.4 | 180 | 22.3 | 29 | 13.7 | 12.5 | 1.45 |
| 56.5 | 120 | 25.0 | 3.00 | 32 | 9.54 | 179 | 24.2 | 29 | 14.8 | 12.6 | 1.19 |
| 58.5 | 80 | 24.6 | 2.50 | 31 | 9.92 | 175 | 22.0 | 26 | 12.4 | 12.9 | 1.42 |

Abbreviations: N = the number of subjects in a lump; Mean = lump mean value; SD = standard deviation.

**Table 2**

Lump mean values of BMI (kg/m$^2$), sit-up score (times/minute), SLJ score (cm), 20-m MSSR score (times), and 10-m SR score (second) for females.

| Age | N | BMI | | sit-up | | SLJ | | 20-m MSSR | | 10-m SR | |
|---|---|---|---|---|---|---|---|---|---|---|---|
| | | Mean | SD | Mean | SD | Mean | SD | Mean | SD | Mean | SD |
| 20.5 | 97 | 21.5 | 3.17 | 33 | 11.7 | 158 | 27.2 | 28 | 13.0 | 12.9 | 1.65 |
| 22.5 | 69 | 22.0 | 3.37 | 30 | 11.6 | 152 | 30.1 | 25 | 12.4 | 13.1 | 1.76 |
| 24.5 | 99 | 22.1 | 3.33 | 32 | 11.4 | 157 | 23.8 | 27 | 13.9 | 12.9 | 1.48 |
| 26.5 | 94 | 22.00 | 3.23 | 31 | 12.1 | 150 | 26.0 | 23 | 12.8 | 13.2 | 1.65 |
| 28.5 | 78 | 22.0 | 3.10 | 28 | 11.7 | 150 | 25.9 | 24 | 11.6 | 13.1 | 1.49 |
| 30.5 | 87 | 22.0 | 3.21 | 30 | 9.04 | 150 | 19.5 | 23 | 8.32 | 13.3 | 1.43 |
| 32.5 | 103 | 22.5 | 3.02 | 26 | 12.7 | 148 | 23.4 | 22 | 11.2 | 13.7 | 1.58 |
| 34.5 | 89 | 22.2 | 2.73 | 26 | 10.4 | 146 | 23.2 | 21 | 10.8 | 13.5 | 1.46 |
| 36.5 | 101 | 22.2 | 2.61 | 26 | 11.8 | 149 | 24.6 | 22 | 10.6 | 13.5 | 1.38 |
| 38.5 | 99 | 22.4 | 2.84 | 23 | 10.1 | 141 | 20.1 | 19 | 8.34 | 13.9 | 1.07 |
| 40.5 | 99 | 23.3 | 3.36 | 27 | 10.4 | 149 | 21.5 | 21 | 10.5 | 13.6 | 1.40 |
| 42.5 | 89 | 22.4 | 3.08 | 24 | 11.9 | 143 | 21.3 | 19 | 7.77 | 13.9 | 1.60 |
| 44.5 | 93 | 22.7 | 2.84 | 26 | 10.1 | 147 | 21.9 | 19 | 9.76 | 13.8 | 1.45 |
| 46.5 | 107 | 23.0 | 2.93 | 22 | 8.67 | 138 | 23.4 | 19 | 9.21 | 14.3 | 1.56 |
| 48.5 | 99 | 23.5 | 3.05 | 23 | 10.3 | 137 | 17.3 | 17 | 7.66 | 14.3 | 1.15 |
| 50.5 | 93 | 23.5 | 2.92 | 22 | 10.3 | 135 | 21.3 | 20 | 11.5 | 14.4 | 1.39 |
| 52.5 | 89 | 23.2 | 2.62 | 21 | 9.86 | 132 | 20.6 | 17 | 8.70 | 14.6 | 1.44 |
| 54.5 | 100 | 23.7 | 2.55 | 20 | 10.5 | 129 | 23.5 | 17 | 7.87 | 14.9 | 1.70 |
| 56.5 | 95 | 23.3 | 2.61 | 18 | 9.84 | 128 | 21.0 | 16 | 8.55 | 14.7 | 1.62 |
| 58.5 | 81 | 23.5 | 2.75 | 18 | 9.79 | 124 | 17.2 | 15 | 6.04 | 15.1 | 1.39 |

Abbreviations: N = the number of subjects in a lump; Mean = lump mean value; SD = standard deviation.

## 3. Results

*3.1. Analysis using the lump mean values*

The LMV of BMI with respect to age fitted to a straight line is shown in Fig. 1, which is expressed as:

$$y(x) = mx + c \qquad (1)$$

where $x$ is age, $c$ is a constant, $m$ is the slope, and $y(x)$ is the LMV of BMI. The parameters $m$ and $c$ are shown in the figure.

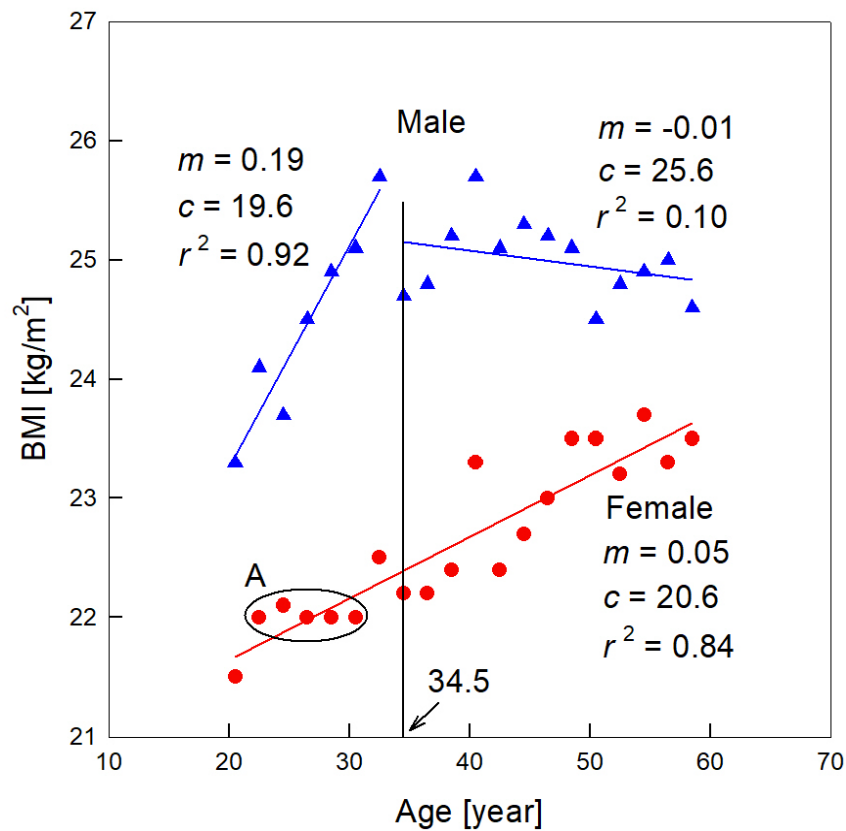

**Figure 1.** Lump mean values of BMI with respect to age.

Figure 1 shows that the females' BMI increases as age increases because $m$ is positive and the goodness of fit for linear regression is sufficient according to the coefficients of determination ($r^2$). For males, the BMI increases more sharply than for females as age increases until the age of 32.5 years with sufficient goodness of fit. In contrast, it decreases slowly after the age of 34.5 years with poor goodness of fit as age increases. This implies that females' BMI and males' BMI before the age of 32.5 years are strongly associated with age, but males' BMI after 34.5 years of age is decreased by different factors. These factors could be concern about health, military service, childbirth, or aging.

It seems that males' BMI increases with age until 32.5 years of age due to changes of life patterns after military service as well as aging, but it decreases with age after that age because they control their weight for health or good body shape. Females' BMI increases with age but is nearly constant in their 20s. Females' BMI seems to increases with age after 32.5 years of age, which is nearly the same as the average age for childbirth in Korea, due to the increase of abdominal obesity after childbirth and aging. But it seems that the females' BMI is nearly invariant in their 20s because they control their weight by controlling food intake for good body shape. To study these results further in relation to fitness, the age dependence of the sit-up test and BMI was investigated (Figs. 2 and 3).

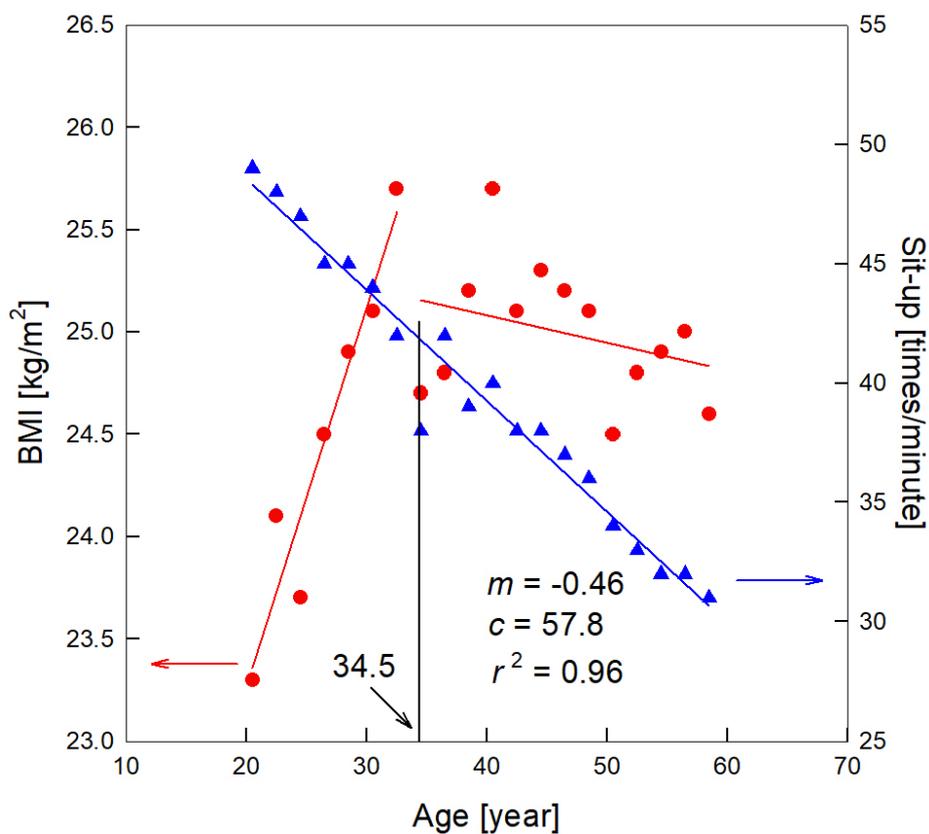

**Figure 2.** Males' BMI and sit-up test scores with respect to age.

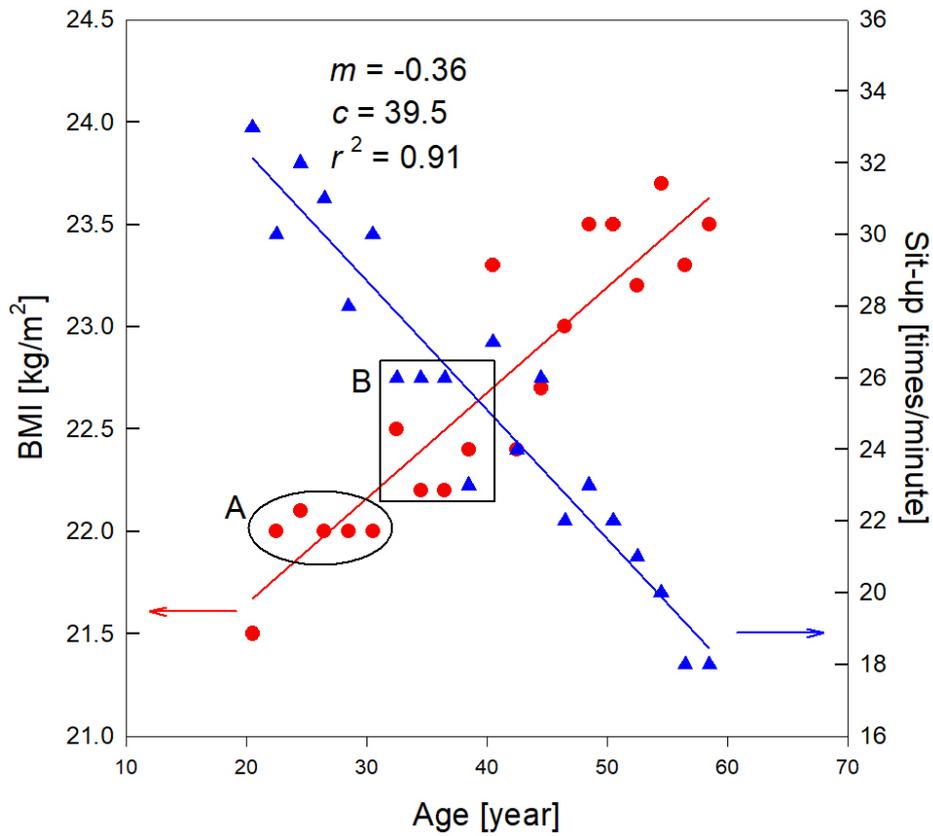

**Figure 3.** Females' BMI and sit-up test scores with respect to age.

Figure 2 shows that the LMV of males' sit-up scores decreases with age and is fitted to the straight line given in Eq. (1), where $y(x)$ is the LMV of sit-up score. The sit-up performance per minute decreases by 0.46 times per year, and the goodness of fit of the linear regressions is sufficient because the coefficient of determination ($r^2$) is sufficiently big. The decrease of the sit-up performance before 32.5 years of age seems to be related to the increase of BMI, but the decrease after 34.5 years of age seems to be affected by other factors because the BMI does not increase after that age. The figure shows that males before 32.5 years of age need to increase their muscular endurance by exercise that reduces their BMI and males after 34.5 years of age need to increase their muscular endurance by exercise that increases muscular strength because their muscular endurance decreases with age independently of BMI.

The LMV of females' sit-up scores with respect to age is shown in Figure 3. The figure shows that the decrease of females' sit-up scores seems to be associated with the increase of BMI. The score is more associated with BMI than that of males' because it decreases with age as the BMI increases. However, there are two important points. One is the plateau in region A, where the sit-up score decreases with age while the BMI is nearly invariant. This can be interpreted that Korean women in their 20s decrease their weight only by controlling food intake without exercise to strengthen their muscular endurance. Therefore, they need to increase their muscular endurance and control their BMI by physical activity or exercise rather than the controlling food intake.

The other point is the irregularity in region B, where the variations of females' BMI and sit-up scores are a little weird. This can be attributed to the body and fitness of women changing abruptly after childbirth. Therefore, they need to increase their muscular endurance and control their BMI by exercise that increases muscular strength and decreases fat.

**Table 3**

$m$ and $c$ for fitting the lump mean values of the BMI and fitness tests ($y$) with respect to age ($x$) to a straight line, $y(x) = mx + c$.

|  | BMI | | Sit-up | | SLJ | | 20-m MSSR | | 10-m SR | |
|---|---|---|---|---|---|---|---|---|---|---|
|  | male | female | male | female | Male | female | male | female | male | female |
| $m$ | 0.02 | 0.05 | -0.46 | -0.36 | -1.16 | -0.78 | -0.63 | -0.29 | 0.06 | 0.06 |
| $c$ | 23.9 | 20.6 | 57.8 | 39.5 | 246 | 174 | 63.4 | 32.0 | 9.17 | 11.7 |
| $r^2$ | 0.19 | 0.84 | 0.96 | 0.91 | 0.97 | 0.89 | 0.97 | 0.90 | 0.97 | 0.94 |

*Note:* The dimension of $c$ is the same as BMI and each fitness test, and the dimension of $m$ is the same as the dimension of BMI or each fitness test divided by year. $r^2$ = coefficients of determination.

The results are summarized in Table 3, including the results for other fitness tests. The table shows that goodness of fit for linear regressions are sufficient for all fitness tests because the coefficients of determination ($r^2$) are sufficiently big. The table shows that the SLJ and 20-m MSSR are negatively associated with the BMI, as in sit-ups, but the 10-m SR is positively associated with the BMI, i.e., the 10-m SR score increases with increasing BMI.

*3.2. Analysis using the fitness sensitivity percentage to age*

Although the goodness of fit for linear regressions are sufficient according to the coefficients of determination (Table 3), the slope, $m$, does not mean the sensitivity of fitness to age because the scale differs according to the type of fitness test, so it cannot be a suitable value to determine the purpose of exercise. In order to determine a suitable exercise for age, the FSPA is introduced:

$$\text{FSPA (Age}_{sub}) \equiv \left| \frac{y_{subA} - y_{best}}{y_{best}} \right| \times 100 \; [\%] \tag{2}$$

where $y_{subA}$ is the fitness test score corresponding to the age of each subject ($\text{Age}_{sub}$), and $y_{best}$ is the best score of each fitness test, which was chosen as the scores of 20.5-year-old males or females.

FSPA means the rate of decrease (sit-up, SLJ, 20-m MSSR) or increase (10-m SR) from $y_{best}$ for each fitness test score. The results for $\text{Age}_{sub} = 40.5$ and 50.5-year-old males and females are shown in Table 4. The sensitivity of fitness to age increases as FSPA increases; i.e., the more FSPA increases, the more the rate of the fitness test score varies. For example, the sit-up score of 40.5-year-old males decreases 19.0% from the best score, $y_{best} = 48.4$ (times/minute), and that of 50.5-year-old-males decreases 28.5% from $y_{best}$; i.e., the rate of the sit-up score decreases more as men grow old. The results are summarized in Table 4, including the results for other fitness tests.

**Table 4**

Fitness sensitivity percentage (FSPA) to age for each fitness test.

|  | Sit-up | | SLJ | | 20-m MSSR | | 10-m SR | |
| --- | --- | --- | --- | --- | --- | --- | --- | --- |
|  | Male | Female | Male | Female | Male | Female | Male | Female |
| $y_{best}$ | 48.4 | 32.1 | 223 | 158 | 50.5 | 26.1 | 10.4 | 12.9 |
| $y_{subA1}$ | 39.2 | 24.9 | 199 | 142 | 37.9 | 20.3 | 11.6 | 14.1 |
| $y_{subA2}$ | 34.6 | 21.3 | 188 | 134 | 31.6 | 17.4 | 12.2 | 14.7 |
| FSPA$_1$ [%] | 19.0 | 22.4 | 10.4 | 9.9 | 25.0 | 22.2 | 11.5 | 9.30 |
| FSPA$_2$ [%] | 28.5 | 33.6 | 15.6 | 14.8 | 37.4 | 33.3 | 17.3 | 14.0 |

*Note:* The subscript 1 denotes the results for 40.5-year-old males and females and the subscript 2 denotes the results for 50.5-year-old males and females. The dimensions of $y_{best}$, $y_{subA1}$, and $y_{subA2}$ are the same as the dimensions of each fitness test.

Table 4 shows that the 20-m MSSR score of males and the sit-up score of females decrease most rapidly with age. For example, the 20-m MSSR score of 40.5-year-old-males decreases 25.0% from the best value, $y_{best} = 50.5$ (times), and the sit-up score of 40.5-year-old females decreases 22.4% from the best value, $y_{best} = 32.1$ (times/minute). This means that it is necessary to decrease the reduction of the 20-m MSSR score (males) and the sit-up score (females) to prevent the degradation of males' cardiovascular endurance and females' muscular endurance with aging. However, it does not mean that reducing these scores is the best way ergonomically to supplement the degradation of fitness by aging; it just means that exercise or physical activity is necessary to increase the 20-m SR score (males) and the sit-up score (females) because their decreasing rates with age are bigger than that of the other fitness scores.

The males' sit-up scores and females' 20-m MSSR scores decrease fast with age too, but the SLJ and 10-m SR scores for males and females decrease the most slowly with age. This means that the types of fitness that decrease most rapidly as age increases are cardiorespiratory endurance and muscular endurance, and the types of fitness that decrease most slowly as age increases are quickness, speed, and agility for males and females.

*3.3. Analysis using the fitness sensitivity percentage to BMI*

In order to establish a goal of exercise according to BMI, the fitness sensitivity percentage to BMI (FSPB) is introduced:

$$\text{FSPB (BMI}_{\text{sub}}) \equiv \left| \frac{y_{\text{subB}} - y_{\text{dB}}}{y_{\text{subB}}} \right| \times 100\ [\%] \tag{3}$$

where $y_{\text{subB}}$ is the fitness test score corresponding to the BMI of each subject (BMI$_{\text{sub}}$) and $y_{\text{dB}}$ is that corresponding to the desired values of BMI (BMI$_{\text{d}}$) for each fitness test, which is chosen as BMIs corresponding to the best scores of each fitness test for convenience in this study (Hwang et al., 2020). But they can be chosen differently according to the desired values of each subject.

FSPB means the rate of each fitness test score to increase (sit-up, SLJ, 20-m MSSR) or decrease (10-m SR) in order to reach BMI$_{\text{d}}$. The results for BMI$_{\text{sub}} = 25.5, 27.5$ (kg/m²) (males) and BMI$_{\text{sub}} = 22.0, 24.2$ (kg/m²) (females) are shown in Table 5. For example, a male with BMI$_{\text{sub}} = 25.5$ (kg/m²) has to increase his sit-up score by 3.2% to attain BMI$_{\text{d}} = $ 23.4 (kg/m²), and a male with $BMI_{sub} = 27.5$ (kg/m²) has to increase his sit-up score by 9.9% to attain the same $BMI_d$.

The sensitivity of BMI to fitness increases as FSPB decreases; i.e., the BMI can be reduced more with decreasing the fitness score when FSPB is smaller. Thus, FSPB corresponds to the width (*b* value) of the Gaussian curve for the fitness test score with respect to BMI (Hwang et al., 2020). *b* means the sensitivity of the BMI to the fitness too; i.e., it means an increase of BMI when the fitness score decreases by half.

**Table 5**

Fitness sensitivity percentage to BMI (FSPB) for each fitness test.

|  | Sit-up | | SLJ | | 20-m MSSR | | 10-m SR | |
| --- | --- | --- | --- | --- | --- | --- | --- | --- |
|  | Male | Female | Male | Female | Male | Female | Male | Female |
| $BMI_d$ | 23.4 | 19.0 | 22.6 | 18.5 | 22.2 | 18.6 | 22.6 | 18.6 |
| $y_{dB}$ | 40.5 | 27.7 | 203 | 148 | 42.3 | 23.8 | 11.4 | 13.5 |
| $y_{subB1}$ | 39.2 | 26.0 | 198 | 144 | 36.9 | 21.2 | 11.7 | 13.8 |
| $y_{subB2}$ | 36.5 | 23.4 | 191 | 138 | 31.6 | 18.3 | 12.0 | 14.2 |
| $FSPW_1$ [%] | 3.2 | 6.1 | 2.7 | 2.9 | 12.8 | 10.9 | 2.6 | 2.2 |
| $FSPW_2$ [%] | 9.9 | 15.5 | 5.8 | 7.0 | 25.3 | 23.1 | 5.3 | 5.2 |

*Note:* The subscript 1 denotes the results for $BMI_{sub} = 25.5$ (males), 22.0 (kg/m²) (females), and the subscript 2 denotes the results for $BMI_{sub} = 27.5$ (males), 24.2 (kg/m²) (females). The dimensions of $y_{dB}$, $y_{subB1}$, and $y_{subB2}$ are the same as the dimensions of each fitness test.

Table 5 shows that the BMI can be controlled effectively with decreasing the 10-m SR score for males and females; e.g., a 2.6% decrease of the 10-m SR score for males with $BMI_{sub} = 25.5$ (kg/m²) and a 2.2% decrease of that for females with $BMI_{sub} = 22.0$ (kg/m²) are necessary to attain $BMI_d = 22.6$ (kg/m²) (males) and $BMI_d = 18.6$ (kg/m²) (females). However, it does not mean that reducing the 10-m SR score is the easiest way kinematically to attain the desired values of BMI; it just means that the desired values of BMI can be attained effectively with decreasing the 10-m SR score.

In contrast, the sensitivity of the BMI to the 20-m MSSR is smallest for males and females; i.e., the reduction rate of the 20-m MSSR score is bigger than those of the other fitness test scores for the same increase of BMI. In other words, increasing the rate of the 20-m MSSR score more is necessary to attain the same desired values of BMI. For example, males with $\text{BMI}_{\text{sub}} = 25.5$ (kg/m$^2$) have to increase their 20-m MSSR score by 12.8% to attain $\text{BMI}_d = 22.2$ (kg/m$^2$), and females with $\text{BMI}_{\text{sub}} = 22.0$ (kg/m$^2$) have to increase their 20-m MSSR score by 10.9% to attain $\text{BMI}_d = 18.6$ (kg/m$^2$).

## 4. Discussion & Conclusion

The BMI of females increased with age, and that of males increased with age until 32.5 years of age, but the BMI of males decreased slightly with age after 34.5 years of age. The degradation of females' fitness and that of males' fitness before 32.5 years of age were attributed to the increase of BMI, but the degradation of males' fitness after 34.5 years of age was attributed to aging with decreases of muscular strength, quickness, cardiopulmonary function, speed, and agility. Males before 32.5 years of age and females of all ages were recommended to increase their fitness by exercise that reduces the BMI. Males after 34.5 years of age needed to increase their fitness by exercise that enhances their fitness rather than reducing the BMI.

Fitness was degraded the most by aging in terms of cardiorespiratory endurance and muscular endurance for males and females. Thus, it was requested to increase them by exercise such as sit-ups and the 20-m MSSR. The sensitivity of the BMI to the 20-m MSSR was smallest, and the BMI could be controlled effectively with decreasing the 10-m SR score. Maintaining fitness after military service and childbirth was important for the health of Korean males and females. Males after 34.5 years of age and females in their 20s should refrain from reducing their BMI by only controlling food intake without exercise to enhance their fitness.

Personal exercise aims can be established using the present results. In Figures 2 and 3, subjects with a score below the straight blue line for the sit-up test have weaker muscular endurance than the mean value of people of the same age. Thus, they need to increase their muscular endurance by exercise that increases the sit-up score. Subjects with a score above the line have strong muscular endurance. The other fitness measures such as quickness, cardiopulmonary function, speed, and agility can be checked in the same manner using the straight line given by Table 3, and exercise to improve weak fitness could be recommended. Although the present results are applicable to Asian adults with similar physique, it is expected that the present method could be applied to investigate the relation between fitness and BMI among other races. It is also expected that the LMV, FSPA, and FSPB could be used to find an effective indicator for fitness and to set personal exercise aims.

**Acknowledgements**

The author would like to thank the Korea Sports Promotion Foundation for providing data from the 2017 Survey of National Physical Fitness. This research did not receive any specific grants from funding agencies in the public, commercial, or non-profit sectors.

**Authors' contributions**

All authors are responsible for the design of the work. NLK performed the data analysis and drafted the manuscript. SCR provided a critical review of the data analysis and the manuscript. All authors read and approved the final version of the manuscript.

**Competing interests**

The authors declare that they have no competing interests.